\documentstyle[11pt,aps,preprint,epsfig,axodraw]{revtex}
\def\gsim{\ \rlap{\raise 3pt \hbox{$>$}}{\lower 3pt \hbox{$\sim$}}\ }
\def\lsim{\ \rlap{\raise 3pt \hbox{$<$}}{\lower 3pt \hbox{$\sim$}}\ }
\newcommand{\bfig}{\begin{center}\begin{picture}}
\newcommand{\efig}[1]{\end{picture}\\{\small #1}\end{center}}
\newcommand{\flin}[2]{\ArrowLine(#1)(#2)}
\newcommand{\wlin}[2]{\DashLine(#1)(#2){3}}

\newcommand{\glin}[3]{\Photon(#1)(#2){2}{#3}}

\newcommand{\sof}{\SetOffset}
\newcommand{\sca}{\SetScale}

\def\Journal#1#2#3#4{{#1} {\bf #2}, #3 (#4)}

\def\NPB{{\em Nucl. Phys.} B}
\def\PLB{{\em Phys. Lett.}  B}
\def\PRL{\em Phys. Rev. Lett.}
\def\PRD{{\em Phys. Rev.} D}
\def\PR{{\em Phys. Rev.} }

\def\ZP{{\em Z. Phys.} }

\def\SPJ{{\em  Sov. Phys. JETP}}
\def\AP{{\em  Ann. Phys.}}
\def\CPC{{\em  Comp.Phys.Comm.}}
\def\JMP{{\em J.Math.Phys.}}

\def\be{\begin{equation}}
\def\ee{\end{equation}}
\def\bea{\begin{eqnarray}}
\def\eea{\end{eqnarray}}

\begin{document}
\preprint{\vbox{\baselineskip=13pt
\rightline{CERN-TH/99-134}
\rightline{hep-ph/9905380}}}

\title{Low-energy neutrino-photon inelastic interactions}

\author{JOAQUIM MATIAS~\footnote{
Talk given at XXXIV$^th$ Rencontres de Moriond: Electroweak Interactions
and
Unified Theories, Les Arcs, France, 13-20 Mar 1999.
}}

\address{TH-Division, CERN, Geneva 23, Switzerland }

\maketitle\

\begin{abstract}
The computation of the polarized amplitudes and cross section of the
processes $\gamma\nu\to\gamma\gamma \nu$,
$\gamma\gamma \to \gamma\nu\bar\nu$ and
$\nu\bar\nu   \to \gamma\gamma\gamma$ is described. We used an effective
lagrangian approach for energies below the threshold for $e^+e^-$
pair production and the complete computation at higher energies for
application in supernova dynamics. Leading contributions of physics beyond
the SM are also commented.
\end{abstract}

\vspace{8.5cm}  

\leftline{May 1999}

\newpage
\section{Introduction}

Low-energy neutrino--photon interactions are of potential interest in
astrophysics, since they could affect the mechanism of
stars to lose energy and hence the study of the stellar evolution, in
particular, supernova dynamics.

However the cross sections of the $2 \rightarrow 2$ processes ($\gamma
\gamma \rightarrow \nu {\bar \nu}$, $\gamma \nu \rightarrow \gamma \nu$
and $\nu {\bar \nu} \rightarrow \gamma \gamma$) are too small. The
reason
for their strong suppression is the prohibition of the coupling of two
photons to a $J=1$ state of any parity. This is simply Yang's
theorem~\cite{yt}. As a consequence the amplitudes of the $2 \rightarrow
2$ processes are zero to order $G_F$. For instance~\cite{dr0},
\bea
A^{\lambda \lambda^{\prime}}(\gamma \nu \rightarrow \gamma \nu)={1 \over 2
\pi} {g^{2} \alpha \over M_{W}^4} \left(1+{4 \over 3} \log {M_W^2 \over
m_e^2} \right) \cos{{\theta \over 2}} f^{\lambda \lambda^{\prime}},
\eea
where $\lambda,\lambda^{\prime}$ are the photon helicities and
$f^{\lambda \lambda^{\prime}}$ is a certain function of  $s$ and $t$.
The important point to notice here is that the scale of the process is
given by $M_W$, which strongly suppresses it at low energies.
However there is a way  to bypass this suppression: if one couples
three photons instead of two the theorem does not apply anymore.
This implies on the one hand, that we will pay an extra $\alpha$ in
the
cross
section. But, on the other, that it is possible
that
the scale of the process is no longer $M_W$ but some other
light scale that could give a large enhancement.

The processes will then be
\bea \label{proc}
\gamma
\gamma \rightarrow \nu {\bar \nu} \gamma,  \qquad \qquad
\gamma \nu \rightarrow \gamma \gamma \nu, \qquad \qquad {\rm and} \qquad \qquad
\nu {\bar \nu} \rightarrow \gamma \gamma \gamma 
\eea

The first may affect the stellar energy loss mechanism and the other
two
may reduce the mean free path of a neutrino inside a supernova core and
in principle they could act as a cut off of high energy photons or
neutrinos by the background. The only existing
computation in the literature of one of these complicated processes
disagrees~\cite{hs}
with a recent computation by Dicus and Repko~\cite{dr1}. Our aim
in~\cite{aqr1}
 was to
settle down this
disagreement and give an explicit derivation of this effective lagrangian
that could be useful for applications in a different context.

Numerically, it was found  that the cross section
of the inelastic 5-leg
processes are  between {\bf  9 and 13 orders of magnitude} larger than
the $2
\rightarrow 2$ corresponding processes for center of mass energies
$\omega$ between 0.2 and 2 times the mass of the electron $m_e$.

However if one needs to go to higher energies, for instance
to energies above
the threshold for $e^+ e^-$ pair production,  an exact calculation
of the processes in Eq. (\ref{proc}) was
important so as to definitively assess their role in astrophysics
and the range of validity of the effective theory.

In this talk we will comment on both approaches: the effective
lagrangian and the direct computation in the SM. We will also briefly
comment on how physics beyond the SM could affect them.

\section{Effective Lagrangian approach}

Three main observations will allow us to obtain  the
leading
contribution at low energies (of the order of $m_e$)
to the cross section of the processes in Eq. (\ref{proc}):

\begin{enumerate}
\item[$ (i)$] First,  the lack of the $1/M_W^4$ suppression in the 5-leg
processes (Yang's theorem does not apply there) suggest that we
concentrate on those diagrams with a lighter particle inside the loop, for
instance $m_e$.

\item[$ (ii)$]  Second, the processes in Eq. (\ref{proc}) are finite, no
counterterm is needed.
This
observation helps in the search for the leading diagrams by
defining a hierarchy of diagrams at low energies. Giving a certain
topology it becomes easier to find its large $m_e$ limit.
These two observations imply that the leading diagrams contributing to
these processes are those given in Fig. 1.
Any other diagram or topology will include other particles inside
the loop, different from the electron, and they will be automatically
suppressed by inverse powers of its mass.

\medskip

\sca{0.8}
\bfig(300,80)
\sof(-40,10)
\flin{40,0}{110,0}  \flin{110,0}{180,0}
\wlin{110,0}{110,20}\
\flin{110,20}{90,40}
\flin{90,40}{110,60}
\flin{110,60}{130,40}
\flin{130,40}{110,20}
\glin{90,40}{65,40}{3}
\glin{110,60}{110,85}{3}
\glin{130,40}{155,40}{3}
\Text(63,-5)[t]{$p_4~~~~\nu$}
\Text(124,-5)[t]{$\nu~~~~p_5$}
\Text(92,12)[l]{$Z$}
\Text(92,56)[l]{$\gamma$}
\Text(61,29)[t]{$\gamma$}
\Text(114,29)[t]{$\gamma$}
\Text(48,32)[r] {$\{i\}$}
\Text(128,32)[l]{$\{k\}$}
\Text(88,72)[b]{$\{j\}$}
\Text(104,44)[r]{$e$}
\Text(130,67)[bl]{$(a)$}
\sof(100,10)
\flin{40,0}{80,0}  \wlin{80,0}{140,0} \flin{140,0}{180,0}
\flin{80,0}{90,20} \flin{90,20}{110,32}
\flin{110,32}{130,20} \flin{130,20}{140,0}
\glin{90,20}{67,37}{3}
\glin{110,32}{110,62}{3}
\glin{130,20}{153,37}{3}
\Text(51,-5)[t]{$p_4~~~~\nu$}
\Text(124,-5)[t]{$\nu~~~~p_5$}
\Text(90,4)[b]{$W$}
\Text(94,40)[l]{$\gamma$}
\Text(55,23)[t]{$\gamma$}
\Text(120,23)[t]{$\gamma$}
\Text(52,36)[r] {$\{i\}$}
\Text(124,36)[l]{$\{k\}$}
\Text(88,56)[b]{$\{j\}$}
\Text(105,26)[r]{$e$}
\Text(130,67)[bl]{$(b)$}
\efig{Figure 1. SM leading diagrams contributing to five-leg
photon--neutrino processes: a) Type A diagrams b) Type B diagrams }

\medskip
\item[$ (iii)$]  So far we have found the leading diagrams. A third
observation will
allow
us to obtain its low-energy expansion in an elegant way:
the subset of diagrams depicted in Fig. 1 resembles a photon--photon
scattering process (see Fig. 2), one of the photons being  off-shell.
\medskip
\end{enumerate}
\medskip


\sca{0.8}
\bfig(300,80)
\sof(40,20)
\glin{110,0}{110,20}{3}
\flin{110,20}{90,40}
\flin{90,40}{110,60}
\flin{110,60}{130,40}
\flin{130,40}{110,20}
\glin{90,40}{65,40}{3}
\glin{110,60}{110,85}{3}
\glin{130,40}{155,40}{3}
\Text(92,10)[l]{$\gamma$}
\Text(106,46)[r]{$e$}
\Text(114,24)[l]{$\gamma$}
\Text(61,28)[t]{$\gamma$}
\Text(92,62)[t]{$\gamma$}
\Text(60,40)[r]{$p_1,~\epsilon_1$}
\Text(114,40)[l]{$p_3,~\epsilon_3$}
\Text(88,72)[b]{$p_2,~\epsilon_2$}
\Text(88,-2)[t]{$p_4,~\epsilon_4$}
\efig{Figure. 2:  Four-photon interaction: Type C diagrams}

\bigskip
\bigskip

Before expressing  this ``resemblance'' in mathematical terms we
should clarify why it is interesting to make this link.
The reason is as follows: we know that the SM is able to describe both the
photon--photon scattering process  and our
inelastic photon--neutrino process which we want to compute. Moreover, we
know that
the low-energy limit of the photon--photon scattering process at one loop
is given by
the Euler--Heisenberg lagrangian~\cite{eh}. Hence if we are able to find a
connection
between the diagrams of the two processes, we will automatically be
able
to establish a link of the effective lagrangian of our process in terms of
the Euler--Heisenberg lagrangian of the photon--photon
scattering without having to compute a single one-loop diagram.

In order to establish the link between those two processes, we should be
able to answer two questions:

\begin{enumerate}
\item[$ (a)$] Can type B diagrams be reduced to type A diagrams?

\item[$ (b)$] Is it possible to relate type A diagrams, which contain a
vertex $Z e e$ with an axial part $i g \gamma_\mu (v_e + a_e \gamma_5)/(2
c_\theta)$, with type C diagrams whose corresponding vertex  $i g
s_\theta \gamma_\mu$ does not?

\end{enumerate}

It was shown in~\cite{aqr1} that it is possible to give a positive
answer
to both questions. The keypoints of the proof are two: on the one hand,
the gauge boson propagators should be expanded in the large
$M_W,M_Z$ limit. This is justified
since we are working at very low energies. On the other hand, we
should find out the
combination of the amplitudes of type A diagrams with different
polarization
indices in such a way that the axial part cancels. Indeed, Gell-Mann
gave an indirect proof in~\cite{ge}, using charge-conjugation
arguments, that such combinations should exist.

In conclusion, at  leading order in $1/M_{W}^{2}$, it was found
in~\cite{aqr1} that the following set
of four diagrams (from a total of 12)  of type A and type B
diagrams is
 proportional to the same integral, called $L_{1}$ (see~\cite{aqr1}
for definitions): \bea \label{set1}
A_{123}^{\alpha \beta \gamma}+A_{321}^{\alpha \beta \gamma}+
B_{123}^{\alpha \beta \gamma}+B_{321}^{\alpha \beta \gamma}=
-{g^{5} s_{W}^3 \over 2} (1+v_{e}) \Gamma_{\mu}
{1 \over {\Delta_{Z} c_W^{2}}} L_{1}^{ \mu \alpha \beta \gamma}\,.
\eea
Similar results are obtained for the other two groups of four diagrams,
the only
difference being a trivial change of momenta and indices inside
$L_{1}^{\mu \alpha\beta\gamma}$.

At this point, the correspondence with the four-photon scattering
 is evident. In fact, by fixing the fourth photon leg
and calling $C_{123}^{\alpha \beta \gamma}$ the corresponding
amplitude in Fig. 2, one finds a contribution proportional
to the same integral $L_1$:
\bea \label{set2}
C_{123}^{\alpha \beta \gamma}+C_{321}^{\alpha \beta \gamma}
=2 {g^4 s_{W}^{4}}
\epsilon_{\mu}({\overrightarrow{P_4}},\lambda_{4})
 L_{1}^{\mu \alpha \beta \gamma}\,,
\eea
and similar results for the two remaining combinations
$C_{132}^{\alpha \beta \gamma}+C_{231}^{\alpha \beta \gamma}$
and
$C_{213}^{\alpha \beta \gamma}+C_{312}^{\alpha \beta \gamma}$.

Up to now, we have proved that both processes are governed by the same
integral, so both should have the same momentum dependence. Moreover, we
have said that at low energies photon--photon scattering is governed by
the
Euler--Heisenberg lagrangian~\cite{eh}.
This automatically tells us which are the only operators that
can be generated at low energies. We have no freedom either in the
structure nor in the relative coefficients between the operators.
Equations (\ref{set1}) and (\ref{set2})
suggest to define a new gauge field
up to the neutrino current
${\tilde A}_{\nu}\equiv{\bar \psi} \gamma_{\nu}(1 -
\gamma_{5}) \psi=2 \Gamma_{\nu}$, with field strength ${\tilde F}_{\mu
\nu}$. Indeed this definition is not unique,  so we should add a
constant in front of the effective lagrangian that still has to be
fixed. The lagrangian then reads
\bea \label{our}
{\cal L}_{\rm eff}={C \over 180}
\left [ 5\left({\tilde F}_{\mu\nu}F^{\mu\nu}     \right)
          \left(       F _{\lambda\rho} F^{\lambda\rho}\right)
-14 {\tilde F}_{\mu\nu}F^{\nu\lambda}F_{\lambda\rho}F^{\rho\mu}\right ].
\eea
In order to fix this constant, we can use the following ratio of
amplitudes
\bea
\lim_{\rm {large~{\it m_e}}}\ \ {{\cal A}_{4 \gamma}^{\rm SM} \over {\cal
A}_{P}^{\rm SM}}=
{{\cal A}_{4 \gamma}^{\rm eff} \over {\cal A}_{P}^{\rm eff}}\,,
\eea
where $P$ is our process and $4 \gamma$ is the photon scattering. Then we
obtain~\cite{aqr1}
\bea
C={g^5 s_{W}^{3} (1 + v_{e}) \over 32 \pi^{2} m_{e}^{4} M_{W}^{2}}=\ {2
G_F
\alpha^{3/2}(1 + v_{e}) \over \sqrt{2\pi} m_e^4}\label{prefactor}\,.
\eea

At this point some remarks are in order. Notice that the suppression
factor for the 5-leg processes is $1/M_W^2$, to be compared with the
$1/M_W^4$ suppression of the  processes with one photon less.
Moreover the scale is not given by $M_W$ but $m_e$, giving rise to an
important enhancement. Finally, once we have established the
dictionary that translates between the two processes,
$
C \rightarrow {\alpha^2 \over m_{e}^{4}}$ and 
$
\Gamma_{\mu} \rightarrow 2
\epsilon_{\mu}({\overrightarrow{P_4}},\lambda_{4})
$,
one can check each polarization amplitude with the corresponding one of
the photon--photon scattering after putting all the photon legs on-shell.
This provides a non-trivial check of our computation.

In~\cite{aqr1} we obtained, using this effective lagrangian, the
eight polarized
differential cross sections for the processes $\gamma \nu \rightarrow
\gamma \nu
\gamma$ and $\gamma \gamma \rightarrow \nu {\bar \nu} \gamma$. Once added
we compared our result with the unpolarized differential cross section and
the total cross section given in~\cite{dr1}. We found perfect
agreement
and hence disagreement with the ones obtained by Hieu and
Shabalin~\cite{hs}.

\section{Direct computation in the SM and beyond}

For energies above the $e^{+} e^{-}$ threshold it is required to
perform a direct computation in the SM~\cite{aqr2,dr2}.
In order to compute the processes in Eq. (\ref{proc}) directly in the SM
we need
to face a multileg computation. The traditional methods, such as 
tensorial decomposition~\cite{passarino},
suffer from large numerical instabilities  due to the
large proliferation of terms and, in particular, due to the appearance
 of gram determinants $\Delta={\rm det} k_i k_j$
 that vanish in collinear regions
of the phase space where the cross section is, indeed,  well defined.
We have used a new method, specially suited for
this type of analysis~\cite{rob}. This is a modified version of the one
of Campbell,
based on grouping the coefficients in sets with a well defined $\Delta
\rightarrow 0$ limit.

\vskip -0.1cm
\vspace{-0.2cm}
\noindent\mbox{\epsfig{file=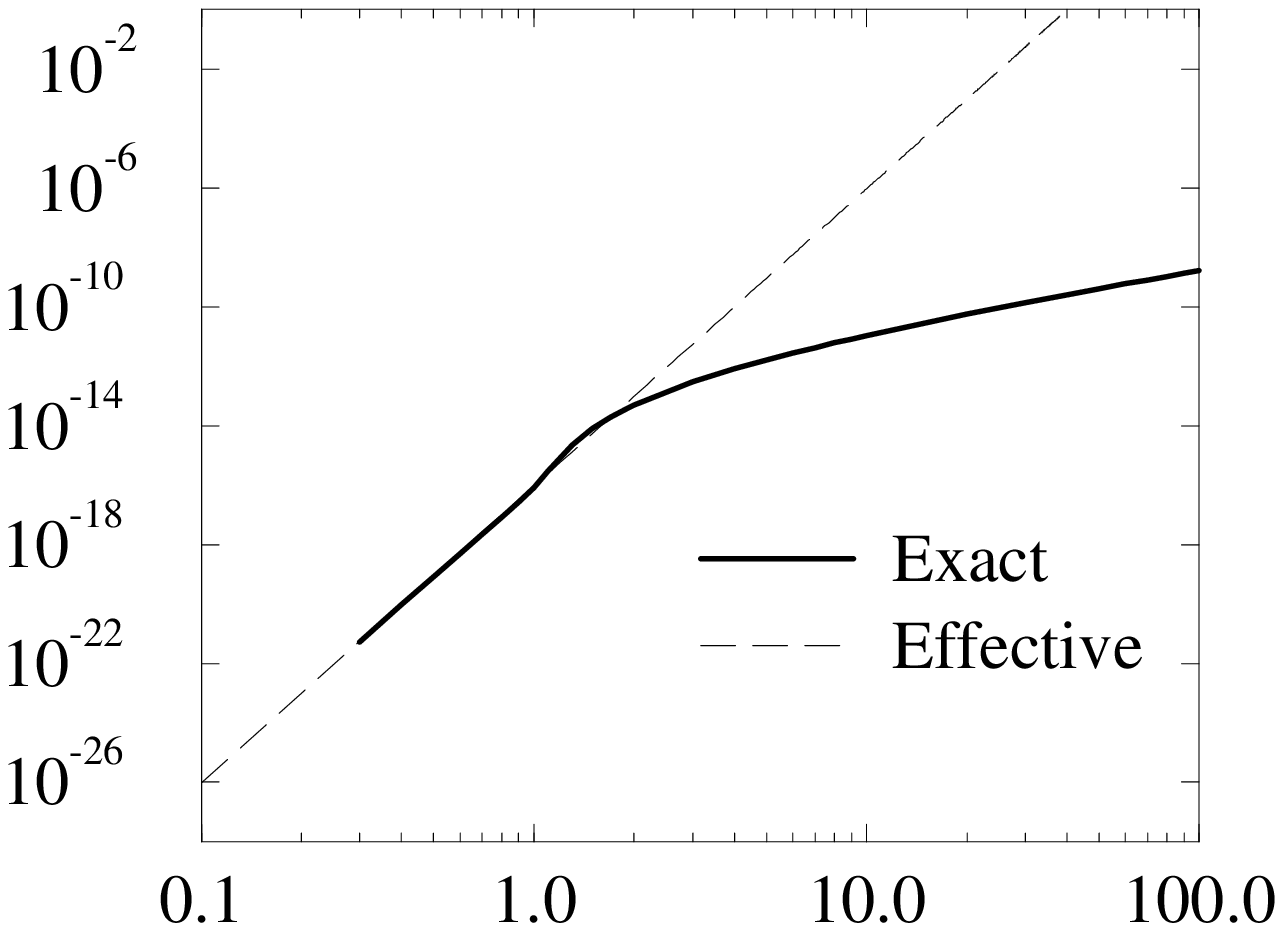,width=3.5in,height=3in}}
\vskip -0.2cm
\vspace{-0.5cm}
\noindent{{{\small{\bf{Figure 3}}}:
{\small{ \hskip -0.1cm$\gamma\nu\to\gamma\gamma\nu$ cross section in fb
as a function of $\omega/m_e$.}}} \protect\label{complet1}
\vspace{0.7cm}

The keypoints of the method are mainly two:
\begin{enumerate}
\item[$ (1)$] The use of a specific representation for the polarization
vectors.
\item[$ (2)$] By means of $\gamma$-algebra and spinor manipulations,
we can reconstruct in the numerators the structures that appear in the
denominators rather than making a tensorial decomposition. In that way all
the results are expressed in terms of
scalar functions with 3 and 4 denominators,
 rank-1 integrals with 3 and 4 denominators,
 rank-2 integrals with 3 denominators,   and
 rank-3 functions with 3 denominators.
This already provides an important simplification with respect
to the standard decomposition, in that the computation of tensors such as
\bea
T^{\mu \nu;\, \mu \nu \rho;\, \mu \nu \rho \sigma} =
\int d^nq \,{q^\mu q^\nu ;\, q^\mu q^\nu q^\rho;\, q^\mu q^\nu q^\rho
q^\sigma \over
D_0\,D_{-1}\,D_2\,D_{(23)}}
\eea
is completely avoided.
\end{enumerate}
 We made a large use of the Kahane--Chisholm
manipulations over
$\gamma$ matrices~\cite{kahane}. Such identities
are strictly four-dimensional, while we are, at the same time,
using dimensional regularization.
Our solution~\cite{aqr2} is splitting,  before any trace manipulation,
the $n$-dimensional integration momentum appearing in the traces
as~\cite{rob} $q\to q +{\tilde{q}}$,
where $q$ and ${\tilde{q}}$ are the four-dimensional and
$\epsilon$-dimensional components ($\epsilon= n-4$), respectively,
so that $q \cdot \tilde{q}= 0$.
The net effects are, on the one hand, that the $\gamma$ algebra can
then be safely performed in four dimensions and, on the other hand,
that a set of  extra integrals
 containing powers of $\tilde{q}^2$ in the numerator arise, but they are
straightforward to compute.

As an example, we have plotted in Fig. 3 the result of the direct
computation for the cross section of the second process~\cite{aqr2} in
Eq. (\ref{proc}). The result is in full agreement with those
reported in~\cite{dr2}. From the plot it is also clear that the
effective theory is valid only when ${\omega} \le 2 m_e$, as expected.
Since the exact formulae are too involved to be given explicitly, we
followed two approaches. We  first tried to extend the
validity of the effective theory by computing the next-to-leading term.
While this was a nice and completely independent check of the results
of the effective theory, we
found that it is not sufficient to enlarge the
range of validity. A second approach was to fit our curves. For instance,
$\sigma(\gamma\nu\to\gamma\gamma\nu)=
\sigma^{\rm eff}(\gamma\nu\to\gamma\gamma\nu)\times
 r^{-2.76046}\times
{\rm exp}\,[2.13317-2.12629 \, {\rm log}^{2}(r)+0.406718 \, {\rm
log}^{3}(r)
-0.029852\, {\rm log}^{4}(r)]\,
$,
where the effective cross section $\sigma^{\rm eff}$ is given in Eq. (26)
of~\cite{aqr2}.
The range of validity is now between $1.7 \le r=\omega/m_e \le 100$.
These fits are useful when these results are used for simulations in
supernovae dynamics.
Finally, in~\cite{aqr3} we studied the leading contribution to the
process
$\gamma \nu \rightarrow \gamma \gamma \nu$ in supersymmetry with R-parity
breaking and in left--right symmetric models (LRSM). The observations
(i) and (ii) in
section 2, also apply here. The leading additional contribution will come
in a LRSM
through the substitution of the $W^{\pm}$ and $Z$ propagators by the
corresponding
$W^{\prime \pm}$ and $Z^{\prime}$, while in the supersymmetric case it
will come through
the
substitution of the $W^{\pm}$ by a slepton. Indeed, it is interesting
to notice that in the  second case the lepton number is no longer
conserved,
i.e. transitions into different neutrino species are allowed. In
particular, we found that in this process
a muon-neutrino prefers to convert into a tau-neutrino rather than an
electron-neutrino, in agreement with SuperKamiokande results. However,
our processes hold for energies much below
the energy of the atmospheric neutrinos ($>1$ GeV) and the cross sections
are still too small. We found that the cross sections in the
$R\!\!\!\!/_p$ MSSM are
enhanced by a factor of the order of few 10$\%$ at best,
relative to the SM cross
sections,
while the correction in LRSM models is negligible.

\section{Astrophysical
implications}\label{conclusion}

While the process
$\gamma\gamma\to\gamma\nu\bar\nu$
provides an energy loss mechanism for stellar process and, in particular,  
it could be
important in the cooling of neutron stars, the other two processes
$\nu\gamma\to\nu\gamma\gamma$ and
$\nu\bar\nu\to\gamma\gamma\gamma$ affect the mean free path  of a
neutrino inside a supernova ans should be included in the supernova codes.

In~\cite{steplitz} it was found, using a Monte Carlo and the results of
the effective theory~\cite{dr1,aqr1} that for a large range of values of
the temperature ($T$) and chemical potential ($\mu$) the mean free path 
was less than
the size of a
supernova core ($10^6$ cm). This result remains valid when using the exact
computation for certain values of $\mu$ and $T$ and at energies not to
far from the $e^+ e^-$ pair production. However, being the exacts results
available it would be of extreme interest to find out the precise values.

Also in~\cite{steplitz} it was predicted using the data of the supernova
SN1987A that the exponent of the energy dependence in the cross section   
$\sigma(\gamma\nu\to\gamma\gamma\nu)\propto  {\omega}^\gamma$ should drop
out from $\gamma=10$ to less than $8.4$ for $\omega$ a few MeV. Using our
results we confirmed
this prediction and indeed, we found that the exponent drops around $3$
for energies between $1$ to $10$ MeV.

Finally, concerning the possible cosmological implications it was
suggested
in~\cite{dr1} that the process $\gamma \nu \rightarrow \gamma \gamma \nu$
might be relevant for some cosmological considerations if the decoupling
temperature~\cite{Peebles} in the exact case is found to be low enough.
With the exacts
results we found~\cite{aqr2} that the temperature is too high to be of any
relevance
in cosmology.

\section*{Acknowledgements}

J.M. acknowledges the financial support from a Marie Curie EC Grant
(TMR-ERBFMBICT 972147).

\end{document}